# AI Climate Emulators Reproduce Their Parents' Temperature Extremes but Lose Land Memory

Gan Zhang[1]

[1]Department of Climate, Meteorology & Atmospheric Sciences, University of Illinois Urbana-Champaign, Urbana, IL, USA

Corresponding author: Gan Zhang (gzhang13@illinois.edu)

**Key Points**

- ACE2 emulators largely reproduce their training data's temperature extremes, including a physical model's tail biases and excess hot days
- ACE2 emulators quickly lose their training datasets' summer land temperature memory, whereas NeuralGCM appears to retain it
- All emulators tested quickly lose the precipitation memory of land, which their training datasets store in soil moisture

**Abstract**

An AI climate emulator imitates its parent, the model or reanalysis it is trained on, so a departure from observations may come from either. Using identical diagnostics of temperature extremes and land memory, we compare three Ai2 Climate Emulators (ACE2) with their parents (SHiELD and ERA5) and with independent physical (CAM6) and hybrid AI (NeuralGCM) models. The ACE2 emulators largely reproduce their parents' temperature extremes. The SHiELD-trained emulators produce SHiELD-like regional tail-bias pattern and excess hot days, while the ERA5-trained emulator matches ERA5's hot days. In summer, the ACE2 emulators' land temperature autocorrelation falls below 0.1 within 8 days, against 12 days in SHiELD and 11 in ERA5. NeuralGCM appears to retain this memory better, but all AI models tested lose most of the precipitation memory that their parents store in soil moisture. Temperature persistence alone thus cannot guarantee the land–atmosphere coupling behind soil-moisture-driven predictability and drought–heat compounding.

**Plain Language Summary**

Some climate applications are embracing AI emulators that learn to imitate a physics-based model and run hundreds of times faster. When an emulator gets extreme temperatures wrong, the remedy depends on whether it inherited the error from the training dataset it learned from or introduced the error itself. We compared three emulators with the datasets they learned from, an independent physics-based model and an observation-constrained dataset. Their extreme-temperature errors are largely inherited. Emulators trained on a physics-based model copy its regional temperature errors and its excess of hot days, whereas the same emulator trained on observation-based data has realistic hot-day counts but adds some regional errors. What these emulators lose is the land's memory. Soil moistened by precipitation keeps cooling the air for weeks, and dry soil keeps heating the air, but the emulators carry no soil moisture. Their summer land temperatures forget past conditions much faster than in the datasets they learned from, although a differently built AI model, NeuralGCM, seems to forget less. However, the three emulators and NeuralGCM lose the land's memory of past precipitation quickly. Emulators can thus look right in temperature while missing a process that matters for droughts, heat waves and weekly-to-monthly forecasts.

## 1 Introduction

AI emulators of global atmospheric models now run stably for decades under prescribed sea surface temperature (SST), at a small fraction of the cost of the physical models they imitate (Chapman et al., 2025; Cresswell-Clay et al., 2025; Kochkov et al., 2024; Watt-Meyer et al., 2025). The computational throughput facilitates the adoption of AI emulators for risk analysis and extreme prediction (e.g., Kent et al., 2025; Mahesh et al., 2025; G. Zhang et al., 2025). Their circulation variability and regional temperature trends have been benchmarked against physical models (e.g., Baxter et al., 2026; Chen et al., 2026; Rucker et al., 2026). Meng et al. (2026) showed that two emulators reproduce the observed frequency of land heat and cold waves, and their persistence, as indicated by autocorrelation, is overall consistent with reference datasets within 5 days. However, an emerging body of studies suggest emulators warm land too little under warmer sea surface temperature (SST) (Henn et al., 2026; B. Zhang & Merlis, 2026), show weak local feedbacks to SST patterns (Van Loon et al., 2025), and give heat-wave projections limited by missing land feedbacks (Duan et al., 2025).

Many existing evaluations often focus on scoring an emulator against observational data. An emulator, however, is trained to imitate one particular model or reanalysis (i.e., its parent), so a departure from observations may be inherited from the parent or created by the emulator. An inherited bias needs a better parent, whereas a created one needs a better emulator. Developers routinely compare an emulator with held-out simulations of its parent (e.g., Watt-Meyer et al., 2025). Some also compare emulator-minus-parent differences beside parent-minus-observation biases in time-mean fields (Duncan et al., 2024; Wu et al., 2026). However, such comparisons are under-utilized for extremes or land interactions. Partitioning a departure requires evaluating the emulator, its parent, and observational data together with the same diagnostics. Additionally, an independent physical model, forced with the same SST, completes the comparison by showing how closely an unrelated model can resemble the parent (i.e., structural uncertainty).

Here we carry out that partition for temperature extremes and land interactions in three emulators of the Ai2 Climate Emulator (ACE) family, trained on two parents (a physical model and a reanalysis). Swapping the parent at fixed architecture shows which biases follow the training data. Changing the architecture, with NeuralGCM (Kochkov et al., 2024) and the models of the AI Model Intercomparison Project (AIMIP) (Henn et al., 2026), shows which biases follow the emulator. Here we emphasize land memory given that most current emulators, unlike physical models and reanalyses, do not explicitly carry soil moisture. Yet soil moisture stores precipitation anomalies for weeks and conditions later temperature and heat extremes (Delworth & Manabe, 1988; Seneviratne et al., 2010; Miralles et al., 2018; Duan et al., 2025). This study seeks to address the following questions:

- Do an emulator's tail biases and hot-day counts come from its parent?
- Without prognostic soil moisture, does it retain its parent's land temperature memory?
- Does it retain the memory of precipitation that soil moisture carries into evaporation?

## 2 Data and Methods

### *2.1 Models and Experiments*

ERA5 (Hersbach et al., 2020) is the parent of the ERA5-trained emulators and also serves as the observational reference. SHiELD, the Geophysical Fluid Dynamics Laboratory's global model (Harris et al., 2020), is the parent of the SHiELD-trained emulators. Two publicly available realizations of SHiELD (~100 km, 1939–2021) are analyzed (Clark et al., 2026). ACE2-SHiELD and ACE2-ERA5 (Watt-Meyer et al., 2025) are deterministic emulators of identical architecture and loss, trained on two SHiELD realizations or on the single ERA5 record over nearly the same years (Text S1). Comparing them swaps the parent but keeps the architecture fixed. ACE2S-SHiELD+ (Clark et al., 2026) is a stochastic successor that adds noise conditioning, a probabilistic loss, multi-step fine-tuning, and a total-energy constraint. Its training used SHiELD's 1979–2008 simulations, as well as slab-ocean and randomly varied $CO_2$ integrations. All three ACE2 variants carry an eight-layer atmospheric column, near-surface fields and land skin temperature, but no soil moisture, soil temperature, snow or canopy water.

Using each of the three ACE2 variants, we conduct 10-member ensemble experiments for 1880–2020. Instead of using the forcings from the emulators' training dataset, we use an independent dataset to avoid favoring any emulators. Specifically, the experiments used daily SST, sea ice, and historical $CO_2$ concentrations that are identical to the Global Ocean–Global Atmosphere (GOGA) experiment with the Community Atmosphere Model version 6 (CAM6) and Land Model version 5 (CLM5) (Danabasoglu et al., 2020). Unlike CAM6, the emulators receive no aerosol, ozone or volcanic forcing, so their forced trends can differ from CAM6's.

Furthermore, SHiELD's own runs used a different observed SST product (Text S1). The deterministic ACE2 members start from the January-1880 states of the ten CAM6 members, and the ACE2S-SHiELD+ members share one state and differ in random seed. More than 70 years of spin-up precede every analysis window.

To test whether the ACE2 results reflect the ACE2 design or emulators in general, we ran ten members of NeuralGCM, which combines a differentiable dynamical core with learned physics (Kochkov et al., 2024). The experiments used its 2.8-degree stochastic-precipitation version (Yuval et al., 2026), under the same GOGA SST for 1989–2019. Like ACE2-ERA5, NeuralGCM is trained on ERA5 (apart from satellite-based precipitation), so comparing the two changes the architecture with the parent fixed. Five-member AIMIP runs of NeuralGCM, NeuralGCM-HRD, ArchesWeather, ArchesWeatherGen (Couairon et al., 2026) and DLESyM (Cresswell-Clay et al., 2025) extend the test to more architectures. AIMIP archives daily output only for 1978–1979 and 2024 (Henn et al., 2026). Given the limited data, each AIMIP model is only classified as showing a deficit, no deficit or an inconclusive result against CAM6 and ACE2-ERA5 subsampled to the same years.

*2.2 Diagnostics*

All diagnostics use daily means over 1990–2019, in June–August (JJA) and December–February (DJF), unless stated otherwise. Tails are the standard deviation, kurtosis and 1st and 99th percentiles of daily 2-m temperature (T2m) anomalies from each source's calendar-day climatology, pooled over land and within 16 land regions (Iturbide et al., 2020). Tail shape is the excess kurtosis of anomalies standardized by each grid cell's calendar-month standard deviation. This metric measures how much of the variability comes from rare, large departures. It is zero for a Gaussian distribution and does not depend on the distribution's width. Hot days are counted at each grid cell as days per year with daily-mean T2m above fixed thresholds of 25 to 35 °C in 2.5 °C steps, then averaged over land.

Extreme-associated circulation states are examined for summer heat waves and winter cold spells in Europe, North America and East Asia (1951–2019). To emphasize societal impacts, we compute the daily share of a region's population (CIESIN, 2018) facing detrended T2m anomalies beyond 1.5 standard deviations. An event is a run of $\geq 3$ days with this share above the 90th percentile for that season based on ERA5 records. Events in each dataset are ranked by the share times the mean anomaly magnitude, summed over the event. For the 14–15 most severe events per region and season, we composite the 500-hPa geopotential height (Z500) anomaly on the peak day and compare composites by pattern correlation over the mapped domain.

Circulation persistence is the 30–70°N Z500 autocorrelation at 5 days, a lag at which it remains well above zero in both seasons. The mean autocorrelation of daily land T2m anomalies at lags of 20, 25 and 30 days represents land temperature memory, r(20–30 d), the time scale of soil moisture memory (Koster & Suarez, 2001; Seneviratne et al., 2006). The excess skin-temperature persistence, the 20–30-day autocorrelation of land skin temperature minus that of T2m, measures how much soil moisture and soil heat storage keep the land surface persistent. Precipitation memory is the correlation between today's land evaporation and precipitation k days earlier, averaged over k = 14, 20, 25 and 30 days.

To find what carries this memory in the parents, we recompute it as a partial correlation, removing the part of today's evaporation explained by same-day soil moisture or by land skin temperature, the emulators' only land state. Memory is measured after removing each member's linear trend and seasonal harmonics, so that a forced trend does not count as memory. This matters for NeuralGCM, which has no $CO_2$ input and warms a third as fast as ERA5 (Text S3). Because all-land averages mix hemispheres with opposite seasons, memory is also reported for Northern Hemisphere (NH) extratropical land (north of 20°N).

A departure from ERA5 is inherited if the parent shares it and created if it does not. Two comparisons provide additional context. The agreement between SHiELD's two members is the upper bound set by internal variability, and CAM6's agreement with SHiELD shows how closely two independent models resemble each other. Memory is compared through the difference (C) and the ratio (R) between an emulator's r(20–30 d) and its parent's. Because NeuralGCM has no 2-m temperature, comparisons involving it use the land-to-ocean ratio of near-surface memory relative to ERA5 (Texts S1 and S3). The 95% confidence intervals are from season-year block bootstraps for memory and event jackknifes for composites (Texts S2 and S3).

## 3 Results

### *3.1 Regional Tail Biases and Hot Days*

Over land, SHiELD's temperature tails are too wide in summer and too narrow in winter (Figure 1a; Table S1). Its JJA 99th-percentile anomaly is 13% larger than ERA5's and its DJF 1st-percentile anomaly 18% smaller in magnitude, several times ERA5's sampling uncertainty. Pooled over the year, the two biases largely cancel. The SHiELD-trained emulators reproduce both biases (ACE2S-SHiELD+, 10% larger and 21% smaller; ACE2-SHiELD, 9% and 17%), whereas ACE2-ERA5 stays within 4% of ERA5. The deterministic emulators also add errors of their own. Within each month, ACE2-SHiELD's tails are heavier than its parent's, and so are ACE2-ERA5's in summer (Table S1).

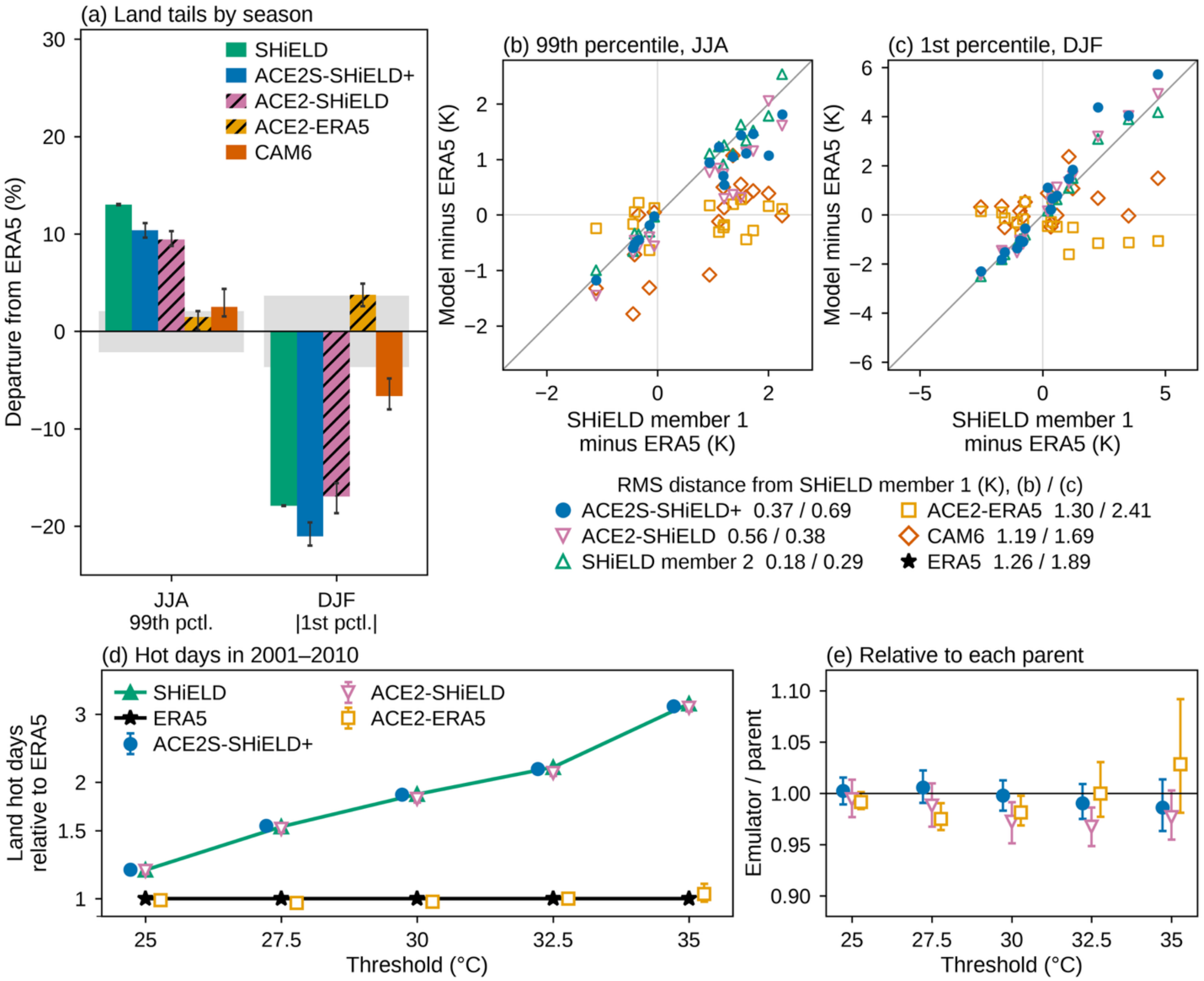

***Figure 1.*** *Regional tail biases and hot-day counts. (a) Departures from ERA5 of the JJA 99th and the DJF 1st percentile of daily T2m anomalies over land without the Greenland and Antarctic ice sheets (area-weighted, 1990–2019). The 1st percentile is compared in magnitude, so a positive departure means a colder cold tail. Bars show member means, whiskers the*

*member range, and grey bands ERA5's 95% sampling interval. (b, c) Departures of the same percentiles from ERA5 in 16 land regions against those of SHiELD member 1, one realization per source (Table S2). Legend values are root-mean-square distances from SHiELD member 1 across the regions; ERA5's is SHiELD's own regional bias. (d) Land days per year above each threshold in 2001–2010, relative to ERA5. The decade was held out of ACE2-SHiELD and ACE2-ERA5 but not of ACE2S-SHiELD+. (e) The same counts relative to each emulator's parent.*

Regional biases of SHiELD and the SHiELD-trained emulators are larger (Figures 1b and 1c; Table S2). Across the 16 examined land regions, SHiELD's JJA 99th and DJF 1st percentiles depart from ERA5's by up to 2.2 and 4.7 K, respectively, while SHiELD's second realization lies only 0.18 and 0.29 K (root mean square across regions) from the first. ACE2S-SHiELD+ and ACE2-SHiELD lie 0.37–0.56 K from SHiELD in summer and 0.38–0.69 K in winter. Swapping the parent, ACE2-ERA5 lies 1.30 and 2.41 K from SHiELD. CAM6, even though it shares forcings with the emulators' experiments, lies 1.19 and 1.69 K away.

Swapping the parent also affects the hot-day bias (Figures 1d and 1e; Table S3). In 2001–2010, held out of both ACE2-SHiELD and ACE2-ERA5, SHiELD produces 1.2–3.2 times ERA5's land days above thresholds of 25–35°C. ACE2-SHiELD carries that excess of hot days, whereas ACE2-ERA5 generally matches ERA5. On its own held-out years, 2012–2019, ACE2S-SHiELD+ matches its parent's counts to within 1.4%. While the hot days in ACE2-SHiELD and ACE2-ERA5 deviate from their parents more at higher thresholds (Figure 1e), the inherited excess mostly results from a bias in SHiELD's mean climate over hot land. SHiELD's typical grid-cell variability over land is comparable to ERA5's (~3 K), and placing ERA5's daily anomalies on SHiELD's calendar-day climatology reproduces 88%–109% of the excess at every threshold (Text S2).

*3.2 Circulation, Autocorrelation, and Land Temperature Memory*

We next check whether circulation contributes to tail and memory biases. For ACE2S-SHiELD+, the composite Z500 patterns of the most severe summer heat events resemble SHiELD's in all three regions (Figures 2a–2i), as do those of winter cold events (Figure S1). Compared directly with the parent's composite, the emulator's pattern correlation (0.41–0.87) is indistinguishable from that of SHiELD's second realization (0.57–0.92) (Figure 2j; Table S4).

CAM6 matches SHiELD's composites similarly, and both match least for East Asian summer heat (Table S4). Because CAM6 resembles SHiELD nearly as well, the composites cannot separate inheritance from a shared SST-forced response, but they reveal no circulation error created by the emulator.

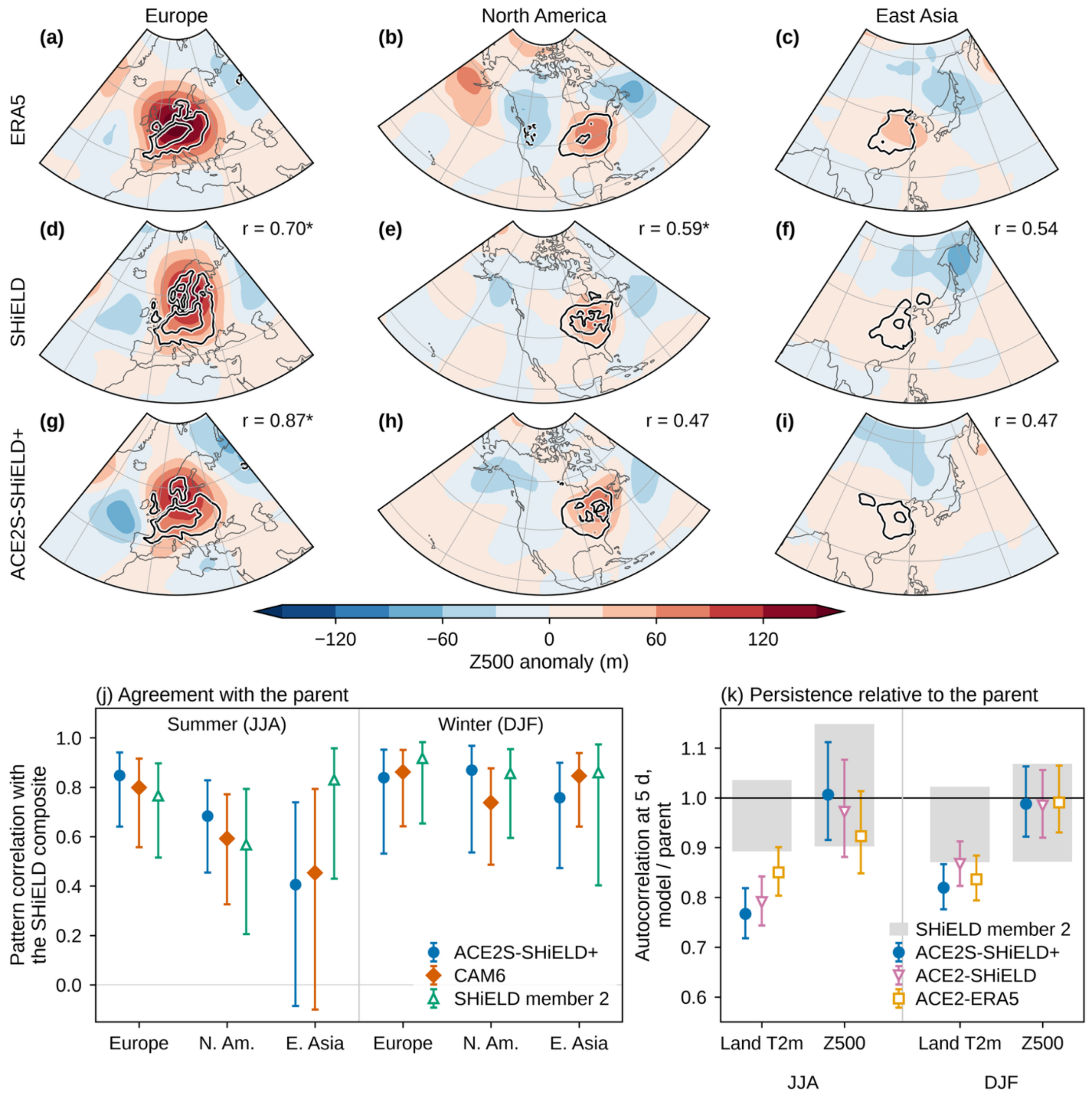

***Figure 2.*** *Circulation during temperature extremes and its persistence. (a–i) Composite Z500 anomaly (shading) and T2m anomaly (contours at −3, 3 and 6 K; negative dashed) on the peak days of the 14–15 most severe summer (JJA) heat events per region (1951–2019). Each map covers the domain of the pattern correlation r with the ERA5 composite, starred where its 95%*

*jackknife interval excludes zero. (j) Pattern correlation with the SHiELD composite for ACE2S-SHiELD+ and CAM6 (mean over SHiELD's two members), and for SHiELD member 2 against member 1 (the attainable upper bound), with 95% jackknife intervals. (k) Autocorrelation at 5 days of land T2m and 30–70°N Z500 in each emulator relative to its parent (1990–2019; detrended anomalies), with 95% bootstrap intervals; grey bands show SHiELD member 2 relative to member 1.*

Circulation persistence is largely captured, though this is not the case for land T2m (Figure 2k). At 5 days the three emulators keep 0.92–1.01 of their parents' Z500 autocorrelation in JJA and 0.98–0.99 in DJF, within the range spanned by SHiELD's two realizations. Their land T2m autocorrelation, however, falls to 0.77–0.85 of the parents' in JJA and 0.82–0.87 in DJF, below that range. The differences suggest that emulators' loss of land T2m memory is not explained by lost circulation persistence and may be attributable to land representation.

The ACE2 emulators' land temperature memory falls below their parents' at every lag beyond a few days (Figures 3a and 3b). In JJA their all-land autocorrelation drops below 0.1 within 8 days, against 12 days in SHiELD and 11 in ERA5. The loss appears in both seasons over NH extratropical land (Table S5). Over NH extratropical land at 20–30 days, ACE2S-SHiELD+, ACE2-SHiELD and ACE2-ERA5 keep 0.28–0.55 of their parents' memory in JJA and 0.36–0.41 in DJF. Over all land, their differences from the parents are C = −0.025 to −0.034 in JJA and −0.036 to −0.054 in DJF, all with 95% intervals excluding zero (Figure 3c). CAM6, forced with the same SST, shows no significant loss against SHiELD in JJA, and in DJF it stays within the difference between SHiELD's two realizations (Figure 3c; Table S6), so the forcing difference does not explain the emulators' loss. The emulators also lack the surface signature of soil memory (Figure 3d; Table S7). In every source with a land surface model, land skin temperature is more persistent than the air above it at 20–30 days. In contrast, the three emulators, whose skin temperature is prognostic but has no soil beneath it, keep at most about half of their parents' excess. Overall, the loss of land temperature memory is not inherited but created by these emulators.

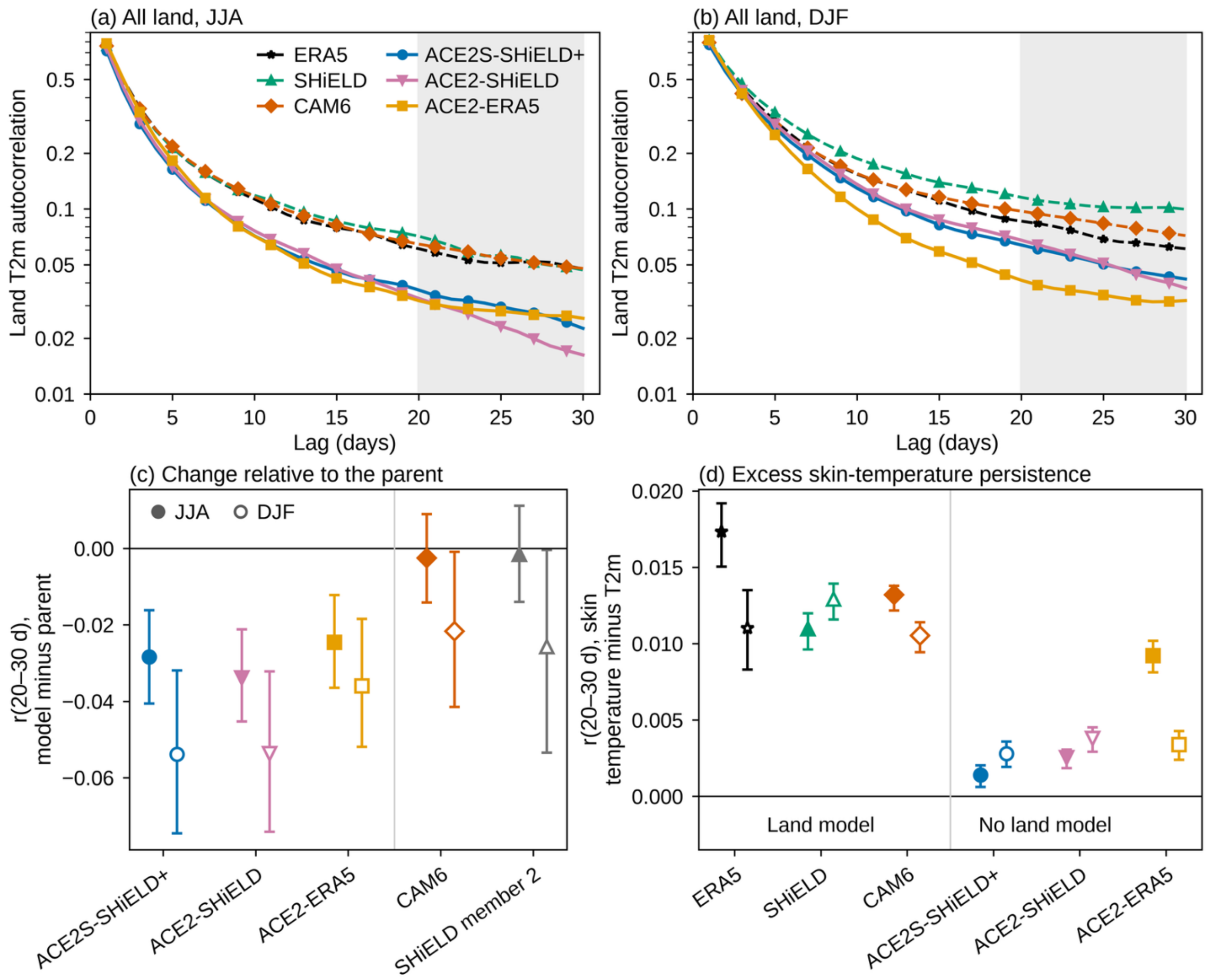

***Figure 3.*** *The ACE2 emulators lose land temperature memory. All panels use the detrended estimator on the common 64 × 128 grid and the data of 1990–2019. (a, b) Lagged autocorrelation of daily land T2m anomalies over all land (logarithmic axis). Shading marks lags of 20–30 days. (c) Difference in r(20–30 d) between each emulator and its parent, between CAM6 and SHiELD, and between SHiELD's two realizations (the parent floor), with 95% bootstrap intervals. (d) The 20–30-day autocorrelation of land skin temperature minus that of T2m, with 95% intervals. In (c) and (d), filled symbols show JJA and open symbols DJF.*

### *3.3 Land Memory Across Architectures*

NeuralGCM, trained on ERA5 like ACE2-ERA5 and likewise without a land state, tests on the same SST whether the loss follows from having no land model (Figure 4a). Because this NeuralGCM version outputs no 2-m temperature, we use its lowest model level as Meng et al. (2026) and compare land-to-ocean memory ratios (Text S1). During summer, NeuralGCM's ratio is ~97% of ERA5's value, resembling its CAM6 and SHiELD counterparts, and significantly higher than ACE2-ERA5's (54%). Part of this contrast reflects NeuralGCM's weaker ocean memory at that height, but its summer memory also exceeds ACE2-ERA5's at 850 hPa, a level all models share (Text S3). Over NH extratropical land alone, the near-surface difference is positive but not significant (Tables S5 and S8). With a fixed climatology, which counts ERA5's faster warming as memory, the ratio contrast is not significant (Table S8). In DJF, NeuralGCM's and ACE2-ERA5's values are indistinguishable (69% and 54%), against the much higher values of CAM6 and SHiELD.

An AIMIP screen extends this test to more architectures, and the outcome again suggests that the loss depends on the model (Text S4). Considering each model's full record (Text S4; Figure S2), ACE2.1-ERA5 keeps 24%–26% of ERA5's summer land memory and ArchesWeatherGen also shows a deficit, whereas NeuralGCM keeps 70%–131%. Lacking a land model thus does not appear sufficient to cause the loss. Furthermore, ACE2S-SHiELD+ is fine-tuned on 5-day rollouts like NeuralGCM (Clark et al., 2026; Yuval et al., 2026) but still loses the memory, so the training horizon alone does not explain the difference. DLESyM is excluded because its SST input, extended over adjacent land, partly prescribes its land surface (Text S4).

Unlike temperature memory, the land's memory of precipitation is lost by every emulator tested (Figures 4b and 4c). ERA5 and physical models suggest soil wetted by precipitation keeps evaporating for weeks. In JJA, the correlation between evaporation and precipitation 14–30 days earlier is 0.043 in ERA5, 0.078 in SHiELD and 0.043 in CAM6. The ACE2 emulators keep only 17%–23% of their parents' value in JJA and 18%–27% in DJF. NeuralGCM, whose evaporation is a column-budget residual, keeps 19% and 16% of ERA5's (Text S3). Summing precipitation over days 14–30 raises these correlations 2.5-fold, but does not fundamentally the emulators' autocorrelation deficit (Text S3). In both parents, the memory passes through the soil (Figure 4d; Table S9). Removing the part of today's evaporation explained by same-day skin temperature leaves it intact or slightly stronger. Removing the part explained by soil moisture removes 92%–93% of it in JJA, as expected when land models compute evaporation from soil moisture.

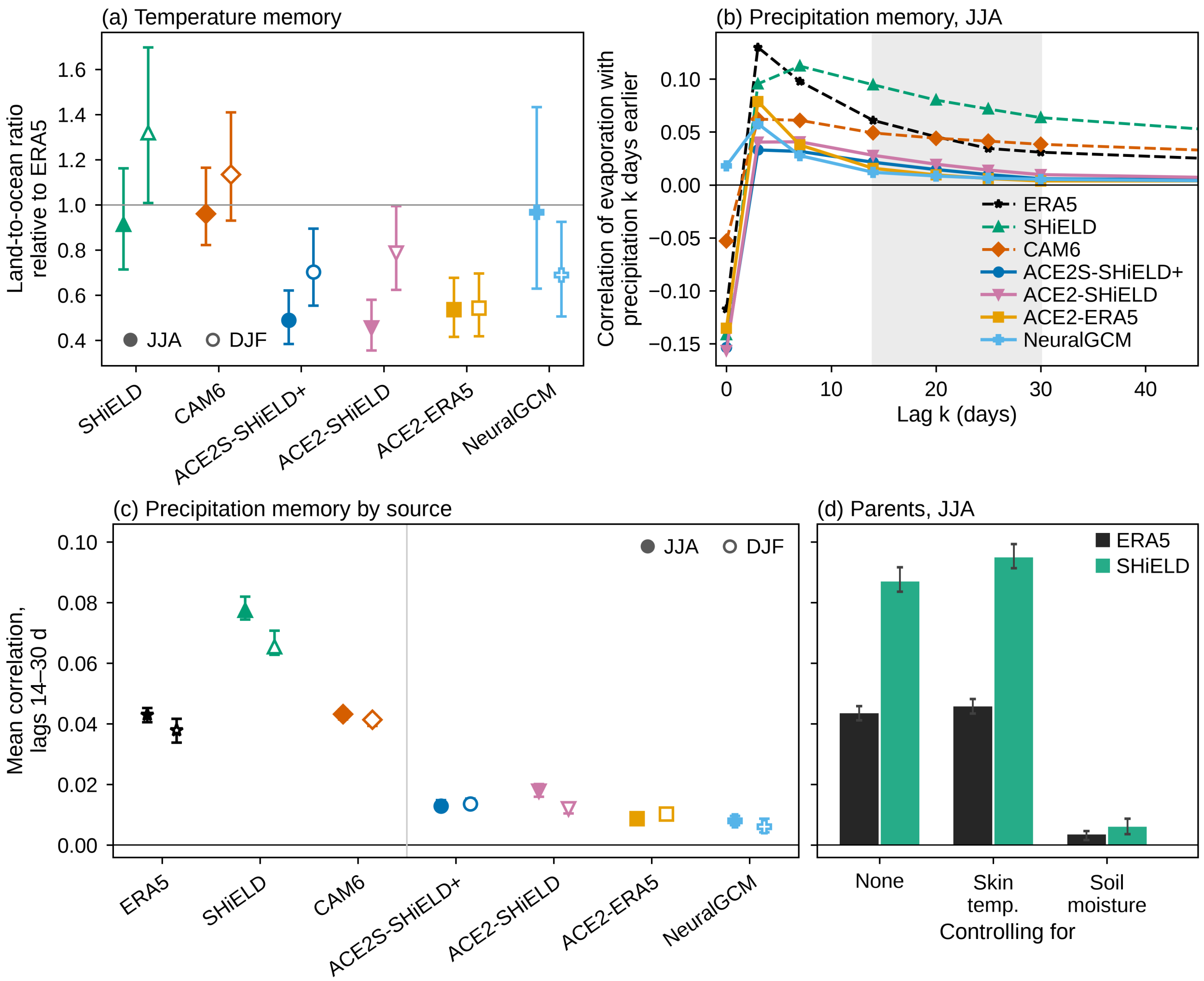

***Figure 4.*** *Temperature memory across model architectures (a) and precipitation memory (b–d). (a) Land-to-ocean ratio of near-surface memory relative to ERA5 in the ten-member GOGA runs (1990–2019, detrended), with 95% confidence intervals. NeuralGCM's near-surface variable is its lowest model level (Text S1). (b) Correlation between land evaporation and precipitation k days earlier, JJA. Shading marks lags of 14–30 days. (c) Its mean at lags of 14, 20, 25 and 30 days, with 95% confidence intervals. In (a) and (c), filled symbols show JJA and open symbols DJF. (d) The same in the parents (JJA, land cells with soil), alone and controlling for same-day skin temperature or four-layer soil moisture.*

## 4 Discussion and Conclusions

Evaluated beside their parents, the ACE2 emulators largely reproduce their parent's temperature-extreme statistics, including the parent's departures from ERA5, but not the land's memory. Regional tail biases and the absolute hot-day excess come from the parent. The parent therefore largely sets these emulators' absolute hot-day statistics, and absolute-threshold statistics from a SHiELD-trained emulator need SHiELD's bias correction. Training on ERA5 does not remove every error, because emulators can create errors of their own.

In contrast, biases in the land's memory are not inherited. The loss of summer land temperature memory appears to be model-dependent. While the ACE2 emulators and ArchesWeatherGen show such biases, NeuralGCM appears to retain ERA5-like summer persistence without a land surface model. However, every emulator we ran loses most of the memory of precipitation. Therefore, temperature persistence alone does not demonstrate realistic land–atmosphere coupling. Because the parents hold the precipitation memory in soil moisture rather than in skin temperature, prognostic soil moisture is likely necessary to restore it, possibly with more than a simple column-total representation (Text S3).

Why the ACE2 emulators lose summer temperature memory while NeuralGCM may not remains open. Neither carries a land state. One hypothesis is that NeuralGCM carries summer persistence in the lower troposphere rather than in the land, as its 850-hPa memory exceeds ACE2-ERA5's over the ocean too (Text S3). The shared winter deficit may instead reflect snow, which none of these models carries. Precipitation memory in evaporation, by contrast, requires a water store. Without that, evaporation cannot depend on precipitation that fell weeks earlier, regardless the architecture of the examined emulators.

These results matter for emulator applications, especially in the transitional zones between wet and dry climates, where soil moisture most strongly controls evaporation and precipitation (Koster et al., 2004). Emulators' appeal for climate risk is large ensembles of extremes, yet such ensembles carry the parent's absolute biases and lack the land's memory of precipitation. Three consequences may follow and warrant further investigation. An emulator without soil moisture probably cannot carry the subseasonal-to-seasonal predictability that initial soil moisture provides (Koster et al., 2006). Additionally, it cannot represent drought–heat compounding, in which a precipitation deficit dries the soil and amplifies later heat (Duan et al., 2025; Miralles et al., 2018). And a land surface that cannot dry out may distort the response to SST warming or other climate forcings (Henn et al., 2026; B. Zhang & Merlis, 2026), possibly leading to underestimated risks of extremes.

Several caveats warrant further discussion. SHiELD and emulator experiments were forced by different SST products. This may account for part of the gap in regional tails but not for the memory loss, as CAM6 does not show it in summer when forced with the same SST. Additionally, some forcings and output variables of the emulators and CAM6 experiments differ. These differences partly arise from developers' design choices and may affect the interpretation of the ACE2-CAM6 comparison. Finally, the AIMIP classification rests on two archived years per member, and ERA5's surface variables are themselves the output of a land model. Therefore, the comparison involves sampling and reference uncertainty. Lastly, the conclusions also apply only to the emulator versions tested, and newer versions need re-evaluation.

For model development, the partition framework is inexpensive and transferable. By evaluating a parent, an emulator, an independent model, and observations together, one may identify whether a bias must be fixed in the parent or in the emulator. The land results may also serve acceptance tests for learned or physical land components coupled to atmospheric emulators. Testing whether prognostic soil moisture also restores the emulators' memory may serve extreme prediction across time scales and is a natural extension for future research.

**Open Research**

Daily ERA5 fields were obtained from the NSF National Center for Atmospheric Research (NCAR) Research Data Archive (ECMWF, 2019). The CESM2 CAM6 prescribed-SST ensembles, whose SST and sea ice also force every emulator and NeuralGCM run, are available from NSF NCAR (Phillips & Simpson, 2020). The SHiELD AMIP output is available from the Allen Institute for Artificial Intelligence (Ai2) on Hugging Face (Ai2, 2026), as are the ACE2-ERA5, ACE2-SHiELD and ACE2S-SHiELD+ checkpoints, which were run with the ACE inference code (Ai2, 2023). NeuralGCM code and checkpoints, including the 2.8° stochastic-precipitation checkpoint used here, are available from its repository (Google Research, 2024), and the AIMIP output from the AIMIP archive described by Henn et al. (2026). The analysis code and derived data supporting this study will be publicly available after peer review.

**Acknowledgments**

G.Z. acknowledges support from the U.S. NSF RISE award 2530555, and computation support by NSF NCAR. G.Z. thanks Ai2 for the ACE2 checkpoints, GFDL/NOAA for SHiELD reference data, ECMWF for ERA5, NCAR for the CESM2 prescribed-SST ensembles, and Google for the NeuralGCM checkpoint. The author thanks Dr. Oliver Watt-Meyer, Dr. Spencer Clark, and other developers for sharing insights about AI emulators. G.Z. used Claude, a generative AI tool, to assist with analysis and writing. The author reviewed all code, results and text and takes full responsibility for the content.

**Conflict of Interest Statement**

G.Z. received computation support of Anthropic's AI for Science Program. G.Z. holds stocks of Alphabet, which is the parent company of Google Research, the developer of NeuralGCM. G.Z. declares no other conflicts of interest for this study.

Supporting Information for

# AI Climate Emulators Reproduce Their Parents' Temperature Extremes but Lose Land Memory

Gan Zhang[1]

[1]Department of Climate, Meteorology & Atmospheric Sciences, University of Illinois Urbana-Champaign, Urbana, IL, USA

## Contents of this file

- Text S1 to S4
- Figures S1 to S2
- Tables S1 to S10

## Introduction

This supporting information documents the experiments, diagnostics and checks behind the main text. Text S1 describes the models, ensembles and forcing; Text S2 the temperature tails, hot days and circulation composites; Text S3 the land-memory estimators and checks; and Text S4 the AIMIP screen. The tables give the values that the main text summarizes, and the figures show the winter composites and the AIMIP screen under every estimator.

**Text S1. Models, Ensembles, and Forcing**

**Training.** ACE2-SHiELD and ACE2-ERA5 (Watt-Meyer et al., 2025) minimize the mean-squared error of two-step (12-hr) forecasts, with corrections that conserve global dry-air mass and close the moisture budget. Both were trained on 1940–1995 and 2011 onward (for ACE2-ERA5, 2011–2019 and 2021–2022) and validated on 1996–2000, which leaves 2001–2010 out of sample; ACE2-SHiELD learned from the two SHiELD realizations analyzed here. ACE2S-SHiELD+ (Clark et al., 2026) was fine-tuned on forecasts of up to 5 days and tested on 2012–2019. For the two ACE2 emulators, 20 of the 30 years in 1990–2019 lie in the training or validation period, so the out-of-sample hot-day test uses 2001–2010 (Figure 1d).

**Forcing.** The GOGA SST and sea ice are based on ERSSTv5, whereas SHiELD's AMIP runs used the CMIP6 AMIP boundary conditions. Besides SST and sea ice, the emulators take only global-mean $CO_2$ and top-of-atmosphere insolation. We built their forcing from CAM6's own fields: SST and sea-ice concentration regridded over ocean only onto the emulators' land mask, an ocean fraction of one minus the land and sea-ice fractions each day, as in Ai2's forcing data, insolation whose monthly global mean matches CAM6's, and CAM6's $CO_2$. Residual differences from CAM6's boundary conditions are confined to coastal cells, where the two grids' coastlines differ. Over 1979–2019 the two SST products agree closely in their monthly anomalies (correlation 0.85) and differ by 0.21 K root-mean-square in the time mean; before the satellite era they differ more. Because CAM6 shares the emulators' forcing and shows no summer memory loss against SHiELD (Table S6), the SST difference cannot explain the emulators' memory loss. It may account for part of the gap between the emulator–parent and parent–parent distances in the regional tails (Text S2).

**NeuralGCM.** We ran the public stochastic-precipitation NeuralGCM (Kochkov et al., 2024; Yuval et al., 2026) with its global-mean surface-pressure filter enabled. It predicts precipitation, learned from IMERG satellite estimates for 2001–2018 rather than from ERA5 (Yuval et al., 2026), diagnoses evaporation from the column water budget, and was trained on forecasts of up to 5 days. Ten members with different random seeds start from an ERA5 state on 1 January 1989; we discard 1989 as spin-up, and all members run stably through 2019. NeuralGCM has no $CO_2$ input and no 2-m temperature, so its lowest model level, about 130 m above the surface, serves as the near-surface variable. All sources are compared on NeuralGCM's grid. Expressing memory as a land-to-ocean ratio relative to ERA5 is meant to absorb the difference in height. At 850 hPa, a level every model provides, NeuralGCM's summer land memory still exceeds ACE2-ERA5's (Text S3).

## Text S2. Temperature Tails, Hot Days, and Circulation

**Tails.** Daily anomalies are departures from each source's calendar-day climatology over 1990–2019, smoothed over ±5 days and not detrended. Figure 1a pools all members over land with area weights, excluding the Greenland and Antarctic ice sheets, on a common 1° grid, separately for JJA and DJF. Tail shape is the excess kurtosis of anomalies standardized by each grid cell's calendar-month standard deviation, which removes the kurtosis that pooling cells or months of different variance would otherwise create. Pooled over the whole year, most of ERA5's tail shape is such mixing. Year-round pooling also hides SHiELD's seasonal biases. Its standard deviation is only 4% below ERA5's over the year but 12% too large in JJA and 11% too small in DJF (Table S1). ACE2S-SHiELD+ reproduces these seasonal departures in sign and most of their size.

***Table S1.*** *Pooled statistics of daily land T2m anomalies as member-mean percentage departures from ERA5, over the whole year and by season (1990–2019). The statistics are calculated on the common 1° grid with area-weighted and without the ice sheets). Shape is the within-month excess kurtosis calculated with anomalies standardized by each cell's calendar-month standard deviation. The ERA5 row gives ERA5's own values. Its 95% season-year bootstrap half-widths are 1.0%–1.2% (s.d.), 2.1%–2.9% (p99) and 3.2%–3.7% (|p1|), and its shape intervals are [0.07, 0.20] (JJA) and [−0.02, 0.06] (DJF).*

| Source | Year-round (s.d. / p99 / \|p1\|) | JJA (s.d. / p99 / \|p1\| / shape) | DJF (s.d. / p99 / \|p1\| / shape) |
|---|---|---|---|
| SHiELD | −4.2 / −6.8 / −9.8 | +12.4 / +13.0 / +14.2 / 0.38 | −11.3 / −12.7 / −17.9 / 0.11 |
| ACE2S-SHiELD+ | −6.3 / −9.0 / −12.0 | +10.1 / +10.4 / +13.0 / 0.40 | −14.1 / −15.1 / −21.0 / 0.12 |
| ACE2-SHiELD | −5.0 / −6.9 / −9.2 | +9.1 / +9.4 / +13.4 / 0.51 | −11.4 / −11.4 / −17.0 / 0.16 |
| ACE2-ERA5 | +2.7 / +3.9 / +3.7 | −0.1 / +1.5 / +0.8 / 0.21 | +4.5 / +5.7 / +3.7 / 0.02 |
| CAM6 | −6.1 / −7.8 / −5.7 | +1.0 / +2.5 / +5.3 / 0.26 | −6.6 / −8.9 / −6.6 / 0.10 |
| ERA5 | 3.49 / 9.66 / 10.29 K | 2.56 / 6.93 / 6.88 K / 0.13 | 4.44 / 12.48 / 13.49 K / 0.02 |

**Regional tails.** Figure 1b–c shows the JJA 99th and the DJF 1st percentile, using one realization per source, pooled without weights within each of the 16 regions (Iturbide et al., 2020). Table S2 here adds the range over members. The SHiELD-trained emulators reproduce SHiELD's regional pattern in every member. For ACE2S-SHiELD+, the correlation with SHiELD's departures is 0.91–0.98 for rank correlations, with area weights within regions, and when any one region is left out, and 0.97–0.99 for the ensemble means, against 0.68 and 0.56 for CAM6's ensemble mean. ACE2-ERA5's departures do not follow SHiELD's, but they are not ERA5's either. Its regional percentiles lie 0.27 and 0.73 K from ERA5's (RMS), closer than CAM6's (0.81 and 0.85 K) but comparable to or beyond the distance between SHiELD's two realizations (0.18 and 0.29 K).

***Table S2.*** *Regional tails across 16 land regions (1990–2019). The table shows the correlation of each source's departure from ERA5 with that of SHiELD member 1, and the root-mean-square distance from ERA5 and from SHiELD member 1, for the JJA 99th and the DJF 1st percentiles. Values are for member 1, as plotted in Figure 1b–c, with the range over members in parentheses.*

| Source | JJA 99th: r with SHiELD | JJA 99th: RMS to ERA5 (K) | JJA 99th: RMS to SHiELD (K) | DJF 1st: r with SHiELD | DJF 1st: RMS to ERA5 (K) | DJF 1st: RMS to SHiELD (K) |
|---|---|---|---|---|---|---|
| SHiELD member 1 (reference) | – | 1.26 | – | – | 1.89 | – |
| SHiELD member 2 | +0.98 | 1.24 | 0.18 | +0.99 | 1.98 | 0.29 |
| ACE2S-SHiELD+ | +0.97 (+0.94 to +0.98) | 1.01 (1.00 to 1.12) | 0.37 (0.27 to 0.44) | +0.98 (+0.98 to +0.99) | 2.39 (2.26 to 2.51) | 0.69 (0.49 to 0.78) |
| ACE2-SHiELD | +0.94 (+0.93 to +0.97) | 0.97 (0.86 to 0.97) | 0.56 (0.53 to 0.62) | +0.99 (+0.96 to +0.99) | 2.14 (1.88 to 2.20) | 0.38 (0.29 to 0.54) |
| ACE2-ERA5 | +0.15 (−0.01 to +0.36) | 0.27 (0.24 to 0.36) | 1.30 (1.21 to 1.36) | −0.64 (−0.71 to −0.36) | 0.73 (0.58 to 0.80) | 2.41 (2.21 to 2.42) |
| CAM6 | +0.71 (+0.65 to +0.71) | 0.81 (0.78 to 0.89) | 1.19 (1.19 to 1.31) | +0.42 (+0.42 to +0.69) | 0.85 (0.81 to 1.30) | 1.69 (1.39 to 1.70) |

**Hot days.** Table S3 gives every threshold of Figure 1d–e and the check of ACE2S-SHiELD+ on its own withheld years. In the mean-state test of Section 3.1, ERA5's anomalies placed on SHiELD's calendar-day climatology reproduce 88%–109% of SHiELD's excess hot days at every threshold, in both realizations, whereas SHiELD's anomalies placed on ERA5's climatology reproduce only −6% to 14%. A uniform warming of ERA5 by SHiELD's land-mean bias (+0.96 K) reproduces only 26% of the excess above 35 °C, because hot days occur where SHiELD's warm bias is largest.

***Table S3.*** *Land days per year with daily-mean T2m above each threshold, as ratios with 95% two-year-block bootstrap intervals. Columns 2–4 cover 2001–2010, withheld from both ACE2 emulators. Column 5 covers 2012–2019, withheld from ACE2S-SHiELD+.*

| **Threshold (°C)** | **SHiELD / ERA5** | **ACE2-SHiELD / SHiELD** | **ACE2-ERA5 / ERA5** | **ACE2S-SHiELD+ / SHiELD, 2012–2019** |
|---|---|---|---|---|
| 25 | 1.19 | 0.994 [0.977, 1.013] | 0.991 [0.985, 1.001] | 1.005 [0.994, 1.017] |
| 27.5 | 1.53 | 0.988 [0.967, 1.010] | 0.975 [0.964, 0.990] | 1.011 [0.995, 1.024] |
| 30 | 1.86 | 0.972 [0.951, 0.991] | 0.981 [0.969, 0.997] | 1.007 [0.987, 1.021] |
| 32.5 | 2.19 | 0.968 [0.948, 0.986] | 1.000 [0.977, 1.030] | 1.006 [0.983, 1.016] |
| 35 | 3.19 | 0.977 [0.955, 1.002] | 1.028 [0.981, 1.092] | 1.014 [0.991, 1.033] |

**Events and composites.** Events follow Section 2.2. The anomalies are standardized by the calendar-day standard deviation in each grid cell, and the population share is computed within each region's detection box (Europe 35–72°N, 15°W–45°E; North America 25–70°N, 170–50°W; East Asia 20–55°N, 100–145°E). The 90th-percentile thresholds come from ERA5 over 1951–2021, and gaps of up to 2 days within an event are merged. The composites use the most severe events through 2019 in ERA5 and in each SHiELD realization and, in the ten-member ensembles, the events that match ERA5's empirical return periods in the pooled members, so every composite samples the same depth of the tail (14–15 events each). Composites are correlated over a domain wider than the detection box (Figure 2). Pattern-correlation intervals come from a delete-one-event jackknife on Fisher z, run over the events of each composite and 95% Student-t limits. We avoid an event bootstrap because duplicated events add noise and bias pattern correlations low. Figure S1 shows the winter composites.

***Table S4.*** *Composite Z500 pattern correlations, 1951–2019, with 95% jackknife intervals. The table shows each model against the ERA5 composite, and ACE2S-SHiELD+, CAM6 and*

*SHiELD member 2 against the SHiELD composite. The last row, ERA5's 1990–2019 composite against its 1951–2019 composite, is context only.*

| Comparison | Europe, JJA | North America, JJA | East Asia, JJA | Europe, DJF | North America, DJF | East Asia, DJF |
|---|---|---|---|---|---|---|
| ACE2S-SHiELD+ vs ERA5 | 0.87 [0.58, 0.96] | 0.47 [−0.03, 0.78] | 0.47 [−0.23, 0.85] | 0.90 [0.52, 0.98] | 0.86 [0.59, 0.96] | 0.74 [0.33, 0.92] |
| SHiELD member 1 / 2 vs ERA5 | 0.70 / 0.82 | 0.59 / 0.42 | 0.54 / 0.62 | 0.89 / 0.85 | 0.82 / 0.84 | 0.77 / 0.65 |
| CAM6 vs ERA5 | 0.91 [0.62, 0.98] | 0.67 [0.35, 0.85] | 0.77 [0.33, 0.93] | 0.95 [0.81, 0.99] | 0.89 [0.67, 0.96] | 0.90 [0.72, 0.97] |
| ACE2S-SHiELD+ vs SHiELD | 0.85 [0.64, 0.94] | 0.68 [0.45, 0.83] | 0.41 [−0.09, 0.74] | 0.84 [0.53, 0.95] | 0.87 [0.54, 0.97] | 0.76 [0.47, 0.90] |
| CAM6 vs SHiELD | 0.80 [0.56, 0.92] | 0.59 [0.33, 0.77] | 0.45 [−0.10, 0.79] | 0.86 [0.64, 0.95] | 0.74 [0.49, 0.88] | 0.85 [0.64, 0.94] |
| SHiELD member 2 vs member 1 | 0.77 [0.52, 0.90] | 0.57 [0.21, 0.79] | 0.83 [0.43, 0.96] | 0.92 [0.65, 0.98] | 0.86 [0.59, 0.95] | 0.86 [0.40, 0.97] |
| ERA5, 1990–2019 vs 1951–2019 | 0.90 | 0.76 | 0.70 | 0.94 | 0.89 | 0.87 |

## Text S3. Land Memory

**Estimators.** The fixed-climatology estimator smooths its calendar-day climatology over ±5 days. The detrended estimator fits four seasonal harmonics with the linear trend and computes an uncentred correlation. The main text uses it for every memory measure, so that a forced trend does not count as memory. This matters most for NeuralGCM, whose summer land warms about a third as fast as ERA5's (0.10 against 0.30 K decade$^{-1}$) because it has no $CO_2$ forcing. Memory intervals are 95% season-year block bootstraps with members resampled. Tables S6 and S8 also give the fixed-climatology estimator.

**Seasons.** All-land DJF mixes Northern Hemisphere winter with summer elsewhere, so Table S5 separates NH extratropical land. In NH winter the SHiELD-trained emulators sit near ERA5's level while SHiELD is much more persistent than ERA5, so they lose much of their parent's winter memory but little relative to observations. The ERA5-trained models lose NH winter memory against ERA5. Removing the trend strengthens the summer loss in all three ACE emulators. In comparison, CAM6 shows no significant summer loss against SHiELD, and its winter difference is no larger than that between SHiELD's own two realizations (Table S6).

***Table S5.*** *Near-surface land memory (1990–2019) at 20–30 days over NH extratropical land, relative to ERA5 and to SHiELD, with the detrended estimator on a common 64 × 128 grid. The values in brackets are 95% intervals. NeuralGCM's values are calculated using its lowest model level, rather than 2-m temperature in other cases. Values relative to SHiELD are omitted for the ERA5-trained models.*

| Source | JJA, vs ERA5 | JJA, vs SHiELD | DJF, vs ERA5 | DJF, vs SHiELD |
|---|---|---|---|---|
| SHiELD | 1.30 [0.84, 2.14] | – | 2.48 [1.57, 4.22] | – |
| CAM6 | 1.35 [0.96, 2.14] | 1.04 [0.77, 1.43] | 1.50 [0.97, 2.49] | 0.60 [0.46, 0.80] |
| ACE2S-SHiELD+ | 0.47 [0.27, 0.82] | 0.36 [0.21, 0.56] | 0.94 [0.60, 1.58] | 0.38 [0.28, 0.51] |
| ACE2-SHiELD | 0.37 [0.21, 0.64] | 0.28 [0.16, 0.44] | 1.01 [0.65, 1.72] | 0.41 [0.30, 0.55] |
| ACE2-ERA5 | 0.55 [0.32, 0.96] | – | 0.36 [0.15, 0.68] | – |
| NeuralGCM | 1.03 [0.45, 2.07] | – | 0.51 [0.21, 0.98] | – |

***Table S6.*** *Land temperature memory (1990–2019) relative to each model's own parent at 20–30 days over land. The difference C with 95% intervals (detrended estimator, as in Figure 3c) and the ratio R under the fixed-climatology and detrended estimators, all on the common 64 × 128 grid. The last row, SHiELD member 2 against member 1, is the parent floor set by internal variability.*

| Model (parent) | C, JJA | R, JJA | R detrended, JJA | C, DJF | R, DJF | R detrended, DJF |
|---|---|---|---|---|---|---|
| ACE2S-SHiELD+ (SHiELD) | −0.028 [−0.041, −0.016] | 0.80 | 0.51 | −0.054 [−0.075, −0.032] | 0.56 | 0.49 |
| ACE2-SHiELD (SHiELD) | −0.034 [−0.045, −0.021] | 0.61 | 0.42 | −0.054 [−0.074, −0.032] | 0.53 | 0.49 |
| ACE2-ERA5 (ERA5) | −0.025 [−0.036, −0.012] | 0.83 | 0.54 | −0.036 [−0.052, −0.018] | 0.77 | 0.50 |
| CAM6 (SHiELD) | −0.003 [−0.014, +0.009] | 1.30 | 0.96 | −0.022 [−0.042, −0.001] | 0.88 | 0.80 |
| SHiELD member 2 (SHiELD) | −0.002 [−0.014, +0.011] | 1.02 | 0.97 | −0.026 [−0.053, −0.000] | 0.77 | 0.76 |

**Precipitation memory.** Table S7 gives the excess skin-temperature persistence (Figure 3d) and the precipitation memory in evaporation (Figure 4c). Precipitation memory is the mean correlation at lags of 14, 20, 25 and 30 days. Its intervals exclude zero by a wide margin for every source, and its small size reflects the noise of single-day precipitation. Correlating evaporation with precipitation summed over days 14–30 raises it 2.1- to 2.6-fold in every source in JJA. The AI models thus still keep only 14%–19% of their parents' value, no more than with single lags. NeuralGCM diagnoses evaporation from the column water budget, and its evaporation correlates positively with same-day precipitation (+0.02, against −0.12 in ERA5). Its precipitation is also learned from IMERG rather than ERA5 (Text S1), so its evaporation-based value does not compare it like for like. We therefore also check precipitation memory in a variable it predicts, the correlation in JJA of near-surface specific humidity with precipitation 14–30 days earlier. ERA5 gives +0.022 [+0.019, +0.025], ACE2-ERA5 +0.008 [+0.007, +0.010] and NeuralGCM +0.011 [+0.009, +0.013]. Both ERA5-trained emulators thus lose much of ERA5's memory in humidity too, keeping 38% and 50% of it, though the fractions are larger than the case of evaporation. Like the near-surface temperature, NeuralGCM's humidity is taken at its lowest model level rather than at 2 m.

***Table S7.*** *Excess skin-temperature persistence (1990–2019) over land at 20–30 days (Figure 3d) and precipitation memory in evaporation, the correlation of land evaporation with precipitation. The values are the average of values at 14, 20, 25 and 30 days earlier (Figure 4c), evaluated with the detrended estimator over land.*

| Source | Excess skin-temperature persistence, JJA | Excess skin-temperature persistence, DJF | Precipitation memory in evaporation, JJA | Precipitation memory in evaporation, DJF |
|---|---|---|---|---|
| ERA5 | +0.0173 [+0.0150, +0.0192] | +0.0110 [+0.0083, +0.0135] | 0.043 [0.041, 0.045] | 0.038 [0.034, 0.042] |
| SHiELD | +0.0110 [+0.0096, +0.0120] | +0.0129 [+0.0116, +0.0139] | 0.078 [0.074, 0.082] | 0.065 [0.063, 0.071] |
| CAM6 | +0.0132 [+0.0122, +0.0138] | +0.0105 [+0.0094, +0.0114] | 0.043 [0.041, 0.045] | 0.041 [0.039, 0.043] |
| ACE2S-SHiELD+ | +0.0014 [+0.0006, +0.0020] | +0.0028 [+0.0019, +0.0036] | 0.013 [0.012, 0.015] | 0.014 [0.013, 0.015] |

| Source | Excess skin-temperature persistence, JJA | Excess skin-temperature persistence, DJF | Precipitation memory in evaporation, JJA | Precipitation memory in evaporation, DJF |
|---|---|---|---|---|
| ACE2-SHiELD | +0.0025 [+0.0018, +0.0031] | +0.0038 [+0.0029, +0.0045] | 0.018 [0.016, 0.020] | 0.012 [0.010, 0.014] |
| ACE2-ERA5 | +0.0092 [+0.0081, +0.0102] | +0.0034 [+0.0024, +0.0043] | 0.009 [0.007, 0.011] | 0.010 [0.009, 0.012] |
| NeuralGCM | n/a (no skin temperature) | n/a (no skin temperature) | 0.008 [0.007, 0.010] | 0.006 [0.004, 0.009] |

**NeuralGCM's temperature memory.** The summer contrast between NeuralGCM and ACE2-ERA5 (Figure 4a) depends on the estimator. It is clear with the detrended estimator but weaker with the fixed-climatology one, which counts the stronger warming of ERA5 and ACE2-ERA5 as memory and so works against NeuralGCM. Repeating both estimators on 1990–2004 and 2005–2019 separately, each with its own climatology, reduces a linear trend's contribution about fourfold (Table S8). The contrast is positive in every half under both estimators, but its interval excludes zero only with the detrended estimator, and with the fixed climatology it vanishes over the full record, so we regard it as likely but not established. Part of the contrast comes from the ocean. Taken separately, NeuralGCM keeps 0.78 [0.48, 1.21] of ERA5's summer land memory and 0.80 [0.72, 0.90] of its ocean memory, against 0.54 [0.41, 0.71] and 1.00 [0.92, 1.11] for ACE2-ERA5, so its lower ocean memory at about 130 m raises its ratio. Over NH extratropical land alone, which needs no ocean reference, NeuralGCM minus ACE2-ERA5 in R is +0.47 [−0.10, +1.35], positive but not significant (Table S8). At 850 hPa, a level every model provides, the summer difference in land R is +0.80 [+0.32, +1.50] with the detrended estimator (+1.82 [+0.36, +7.13] over NH extratropical land) and +0.31 [−0.05, +0.87] with the fixed climatology. NeuralGCM is also more persistent than ACE2-ERA5 over the ocean at that level (+0.44 [+0.24, +0.67]), so its summer persistence is not confined to land. In DJF, no land difference is significant at either level with the detrended estimator, and the land-to-ocean ratios are 0.69 [0.51, 0.92] for NeuralGCM and 0.54 [0.42, 0.70] for ACE2-ERA5.

***Table S8.*** *Summer (JJA) land-to-ocean near-surface memory ratio relative to ERA5, on the full record and on each 15-year half, under both estimators computed within each window. The NeuralGCM-minus-ACE2-ERA5 difference in that ratio and in R over NH extratropical land.*

| **Window** | **Estimator** | **NeuralGCM** | **ACE2-ERA5** | **CAM6** | **NeuralGCM minus ACE2-ERA5** | **Same, NH extratropical land R** |
|---|---|---|---|---|---|---|
| 1990–2019 | Fixed climatology | 0.75 [0.50, 1.15] | 0.78 [0.64, 0.97] | 0.88 [0.75, 1.08] | −0.03 [−0.29, +0.32] | −0.01 [−0.45, +0.63] |
| 1990–2019 | Detrended | 0.97 [0.63, 1.43] | 0.54 [0.42, 0.68] | 0.96 [0.82, 1.16] | +0.43 [+0.11, +0.86] | +0.47 [−0.10, +1.35] |
| 1990–2004 | Fixed climatology | 0.60 [0.44, 0.79] | 0.47 [0.32, 0.63] | 0.86 [0.68, 1.12] | +0.13 [−0.02, +0.31] | +0.03 [−0.22, +0.31] |
| 1990–2004 | Detrended | 0.60 [0.45, 0.82] | 0.35 [0.21, 0.50] | 0.79 [0.63, 1.01] | +0.25 [+0.08, +0.47] | +0.25 [−0.01, +0.63] |
| 2005–2019 | Fixed climatology | 1.17 [0.63, 2.08] | 0.84 [0.65, 1.13] | 1.09 [0.91, 1.47] | +0.32 [−0.25, +1.14] | +0.65 [−0.38, +2.86] |
| 2005–2019 | Detrended | 1.41 [0.71, 2.66] | 0.57 [0.35, 0.94] | 1.25 [0.96, 1.97] | +0.84 [+0.17, +1.96] | +1.28 [−0.20, +6.87] |

**Where the parents keep precipitation memory.** For SHiELD and ERA5, we recompute the precipitation memory in each grid cell as a partial correlation, controlling for the same day's moisture in the four soil layers (0–10, 10–40, 40–100 and 100–200 cm in SHiELD; 0–7, 7–28, 28–100 and 100–289 cm in ERA5), for column soil water alone, or for land skin temperature (Table S9; Figure 4d). Cells without soil, mainly the ice sheets, are dropped, leaving 90% (SHiELD) and 98% (ERA5) of the land area. Both land models compute evaporation from soil moisture, so a large removal is expected. The four soil layers remove 78%–93% of the memory in both parents and seasons, and at least 71% over NH extratropical land, so atmospheric persistence carries little of it. Column soil water alone removes only 30%–46% in JJA and 36%–51% in DJF. Because the column-total representation mixes fast-draining shallow layers with slow deep ones, so adding it to an emulator would carry only part of this memory. Skin temperature removes none (−9% to +2%).

***Table S9.*** *Precipitation memory in the two parents (Figure 4d). The mean over lags 14–30 days of the correlation between land evaporation and earlier precipitation, without and with same-day controls. The calculation is conducted with the detrended estimator for land cells with soil (1990–2019), with 95% intervals and the share removed by each control. Dropping cells without soil raises SHiELD's "no control" value above that of Table S7.*

| Source | Controlling for | JJA | DJF |
|---|---|---|---|
| SHiELD | No control | 0.087 [0.084, 0.092] | 0.073 [0.070, 0.079] |
| SHiELD | Four soil layers | 0.006 [0.004, 0.009]; 93% removed | 0.016 [0.014, 0.019]; 78% removed |
| SHiELD | Column soil water | 0.047 [0.044, 0.051]; 46% removed | 0.036 [0.034, 0.040]; 51% removed |
| SHiELD | Land skin temperature | 0.095 [0.091, 0.099]; −9% removed | 0.076 [0.073, 0.081]; −4% removed |
| ERA5 | No control | 0.044 [0.041, 0.046] | 0.039 [0.034, 0.042] |
| ERA5 | Four soil layers | 0.004 [0.002, 0.005]; 92% removed | 0.004 [0.001, 0.006]; 90% removed |
| ERA5 | Column soil water | 0.031 [0.028, 0.032]; 30% removed | 0.025 [0.021, 0.027]; 36% removed |
| ERA5 | Land skin temperature | 0.046 [0.043, 0.048]; −5% removed | 0.038 [0.034, 0.041]; 2% removed |

## Text S4. The AIMIP Screen

**Data and classification.** AIMIP archives daily near-surface temperature only for 1978–1979 and 2024 (DLESyM: 1978–1984), which gives two summers and two winters per member (Henn et al., 2026). The models are trained on ERA5 for 1979–2014, so 1979 lies inside their training period and 2024 outside it. AIMIP archives no daily evaporation, so precipitation memory is assessed only in our GOGA runs. Land memory, the 20–30-day autocorrelation of daily land near-surface temperature anomalies, is expressed relative to ERA5's 1990–2019 value under the same estimator. We calibrate the classification with 1,000 samples (300 for Xm and Xp) of AIMIP shape drawn from our 30-year records. A model shows a deficit if its value falls below the 5th percentile of the CAM6 samples, no deficit if it exceeds the 95th percentile of the ACE2-ERA5 samples, and an inconclusive result otherwise.

**Estimators.** With two short windows per member, the result depends on how the anomalies are defined, so we report four estimators. The first estimator (U) fits the seasonal cycle and a mean for each member and window to the AIMIP windows themselves. With only two seasons, subtracting the sample mean lowers the 20–30-day correlation by about as much as the signal, so U places almost every model near or below zero. The second estimator (X) and its variants avoid this bias and are thus adopted by the main text. This estimator takes anomalies from a climatology (a linear trend plus four seasonal harmonics) fitted to each model's full monthly record for 1978–2024, pooled over members, and computes an uncentred correlation. Because some NeuralGCM members run persistently warmer or cooler than others, two variants subtract each member's mean anomaly (Xm) or fit each member its own climatology (Xp). Xp corresponds to the detrended estimator of the GOGA runs (Text S3). DLESyM is excluded because AIMIP extends its SST input over adjacent land, which partly prescribes its land surface; we made this decision after seeing its values.

***Table S10.*** *AIMIP land memory in JJA and DJF relative to ERA5's 1990–2019 value, with the classification, under two estimators. U is biased low on records this short, so the main text uses X and its variants (Figure S2).*

| **Model** | **U, JJA** | **U, DJF** | **X, JJA** | **X, DJF** |
|---|---|---|---|---|
| ACE2.1-ERA5 | −0.14 (inconclusive) | −0.30 (inconclusive) | 0.25 (deficit) | 0.67 (deficit) |
| NeuralGCM | −0.02 (inconclusive) | −0.27 (inconclusive) | 1.31 (no deficit) | 1.05 (no deficit) |
| NeuralGCM-HRD | −0.03 (inconclusive) | −0.28 (inconclusive) | 1.33 (no deficit) | 1.08 (no deficit) |
| ArchesWeather | 0.02 (inconclusive) | −0.17 (inconclusive) | 0.71 (inconclusive) | 1.04 (no deficit) |
| ArchesWeatherGen | −0.03 (inconclusive) | −0.42 (deficit) | 0.57 (deficit) | 0.64 (deficit) |
| DLESyM (excluded) | 0.68 (no deficit) | 0.97 (no deficit) | 1.46 (no deficit) | 1.45 (no deficit) |

## Supporting Figures

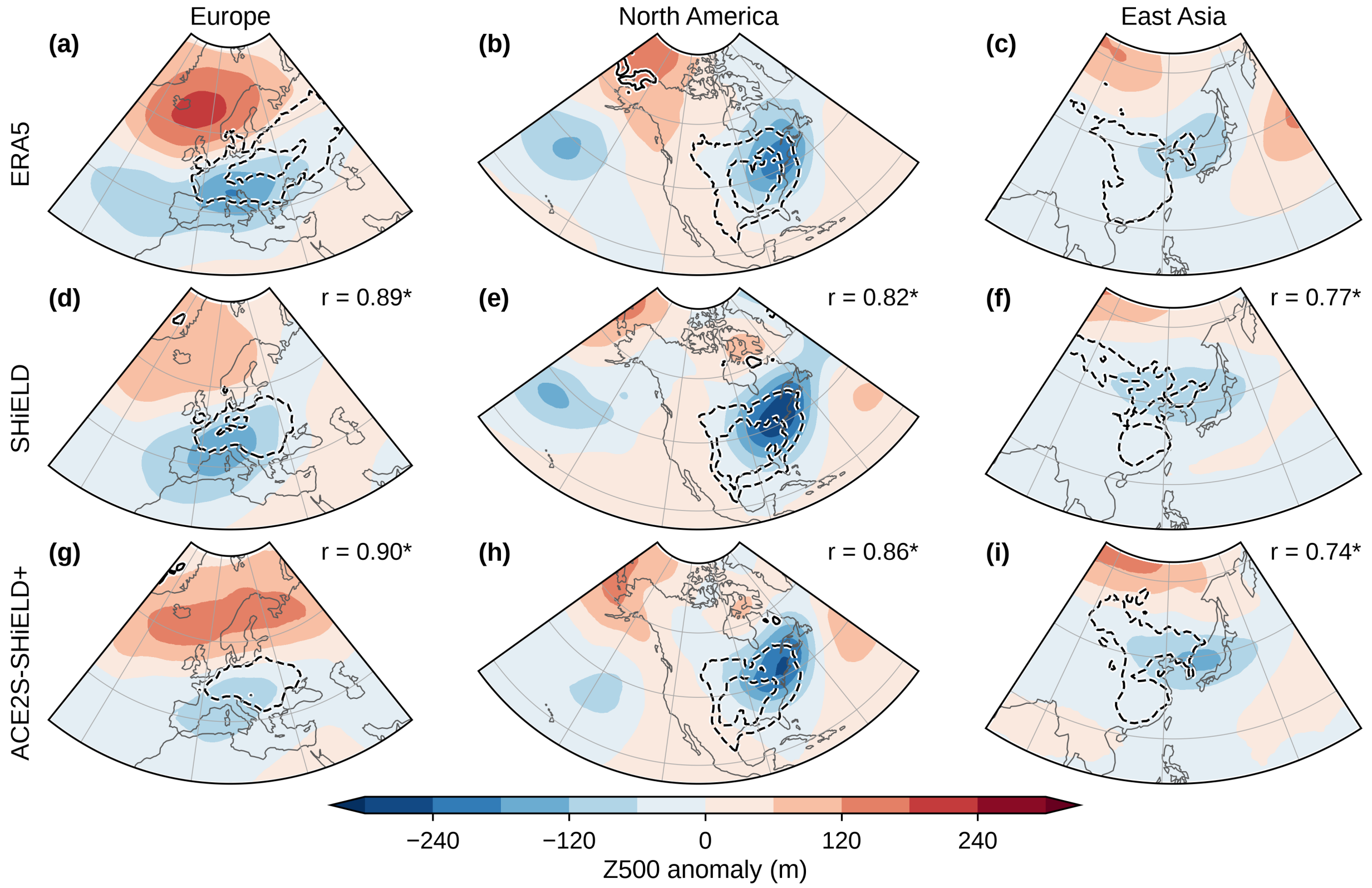

***Figure S1.*** *As Figure 2a–i, for the 14–15 most severe winter (DJF) cold events, 1951–2019. T2m contours are at −12, −6 and 6 K (negative dashed).*

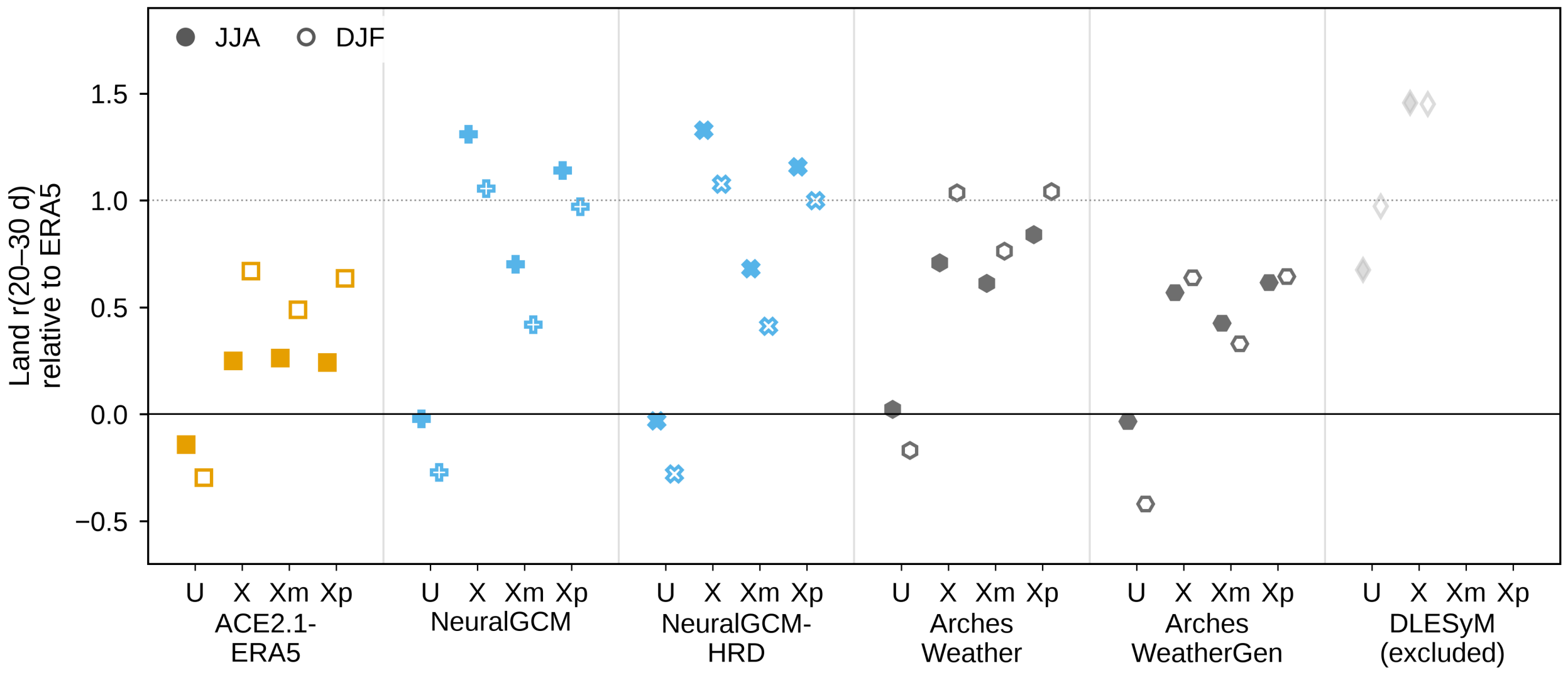

***Figure S2.*** *AIMIP land memory relative to ERA5, as in Table S10, under all four estimators of Text S4. Filled symbols show JJA and open symbols DJF.*